\documentclass[amsmath,amssymb,longbibliography,superscriptaddress]{revtex4-2} %

\usepackage[T1]{fontenc}
\usepackage[utf8]{inputenc}
\usepackage{graphicx}
\usepackage{amsmath}
\usepackage{amssymb}
\usepackage{xcolor}
\usepackage{tablefootnote}
\usepackage{tabularx}
\usepackage{upgreek}
\usepackage{siunitx}
\usepackage{bm}
\usepackage{cleveref}  

\newcommand{\ie}{{i.\,e. }}

\newcommand{\beq}{\begin{equation}}
\newcommand{\eeq}{\end{equation}}
\newcommand{\dd}{\mathrm{d}}

\newcommand{\de}{\Delta\varepsilon}
\newcommand{\es}{\varepsilon_\mathrm{s}}
\newcommand{\ei}{\varepsilon_\infty}
\newcommand{\ez}{\varepsilon_0}
\newcommand{\ep}{\varepsilon^\prime}
\newcommand{\edp}{\varepsilon^{\prime\prime}}
\newcommand{\cdp}{\chi^{\prime\prime}}
\newcommand{\gk}{g_\mathrm{K}}
\newcommand{\bmu}{\bm{\mu}}

\crefname{figure}{Fig.}{Figs.}

\begin{document}

\title{On the Spectral Shape of the Structural Relaxation in Deeply Supercooled Liquids}

\author{Till Böhmer}
\affiliation{Glass and Time, IMFUFA, Department of Science and Environment, Roskilde University, 4000 Roskilde, Denmark.}
\email{tillb@ruc.dk}
\author{Florian Pabst}
\affiliation{SISSA—Scuola Internazionale Superiore di Studi Avanzati, 34136 Trieste, Italy}
\email{fpabst@sissa.it}
\author{Jan Gabriel}
\affiliation{Institute of Materials Physics in Space, German Aerospace Center, 51170 Colone, Germany}
\author{Rolf Zeißler}
\affiliation{Institute for Condensed Matter Physics, Technical University of Darmstadt, 64289 Darmstadt, Germany.}
\author{Thomas Blochowicz}
\affiliation{Institute for Condensed Matter Physics, Technical University of Darmstadt, 64289 Darmstadt, Germany.}
\email{thomas.blochowicz@physik.tu-darmstadt.de}

\date{\today} 

\setlength\parindent{0pt}

\newpage

\begin{abstract}
Structural relaxation in deeply supercooled liquids is non-exponential. In susceptibility representation, $\cdp(\nu)$, the spectral shape of the structural relaxation is observed as an asymmetrically broadened peak with a $\nu^{1}$ low- and $\nu^{-\beta}$ high-frequency behavior. In this perspective article we discuss common notions, recent results and open questions regarding the spectral shape of the structural relaxation. In particular, we focus on the observation that a high-frequency behavior of $\nu^{-1/2}$ appears to be a generic feature in a broad range of different deeply supercooled liquids. Moreover, we review extensive evidence that contributions from orientational cross-correlations can lead to deviations from the generic spectral shape in certain substances, in particular in dielectric loss spectra. Additionally, intramolecular dynamics can contribute significantly to the spectral shape in substances containing more complex and flexible molecules. Finally, we discuss the open questions regarding potential physical origins of the generic $\nu^{-1/2}$ behavior and the evolution of the spectral shape towards higher temperatures.
\end{abstract}

\newpage
\maketitle 
\section{Introduction}
Already  170 years ago in his work on the Leyden jar Rudolf Kohlrausch demonstrated a non-exponential relaxation of the electric charge in a cylinder-shaped capacitor\cite{Kohlrausch1854}. Today, dielectric spectroscopy is routinely used to monitor the relaxation of permanent molecular dipole moments in all kinds of materials, in particular in supercooled liquids and glasses. While the early \emph{Debye}-model of dipolar relaxation suggested an exponential decay of dipolar correlations \cite{Debye1913} it soon became clear that a single relaxation time $\tau$ is the exception rather than the rule, in particular in supercooled liquids. In this regard, the term \textit{relaxation stretching} is used to indicate that structural relaxation in supercooled liquids is considerably stretched along the time-axis compared to an exponential decay \cite{Ediger1996,Cavagna2009}.

Most often, relaxation stretching in supercooled liquids is  discussed in terms of a relaxation time distribution $G(\ln\tau)$, so that, for a complex susceptibility $\hat\chi(\omega)$ that may be derived, e.g., from a dielectric or also a depolarized  light scattering experiment one can write: 
\begin{equation}
    \hat\chi(\nu)=\Delta \chi \int\limits_{-\infty}^{\infty}G(\ln\tau)\frac{1}{1+ i 2\pi\nu\tau}~\mathrm{d}\ln\tau.
    \label{equ:glntau}
\end{equation}
Or, equivalently, in the time domain for a time autocorrelation function $C(t)$:
\begin{equation}
    C(t)\propto \int\limits_{-\infty}^{\infty}G(\ln\tau)\ \exp{(-t/\tau)}~\mathrm{d}\ln\tau.
    \label{equ:glntau-t}
\end{equation}
Formally, in equilibrium and in the linear response regime, any relaxation function can  be decomposed into a superposition of exponential decays and thus, a relaxation time distribution can be obtained \cite{Hansen1976a}. However, the physical meaning of $G(\ln\tau)$ and whether there are indeed slow and fast relaxing molecules in the supercooled liquid at any given time, is an entirely different question. 

The universal observation for the structural relaxation or $\alpha$-process in supercooled liquids is that the associated loss peak in $\chi^{\prime\prime}(\nu)$, the imaginary part of $\hat\chi(\nu)$, is asymmetrically broadened with a high-frequency power law $\propto \nu^{-\beta}$, where $\beta<1$, and a low-frequency power law $\propto\nu$. From the viewpoint of Eq.~\ref{equ:glntau}, the latter fact implies the distribution of relaxation times to be cut-off at large $\tau$ and, thus, there exists a slowest relaxation mode in supercooled molecular liquids. The origin of the asymmetrical broadening of the relaxation peak and its implications for the physics of glassy materials has been a matter of debate for several decades - yet a widely accepted consensus remains elusive.

This perspective article discusses recent developments regarding the spectral shape of the structural relaxation in deeply supercooled liquids. By "deeply supercooled", we refer to temperatures well below the critical temperature predicted by the mode-coupling theory, where dynamic heterogeneity is widely considered to play a crucial role. The article is motivated by a growing evidence that the spectral shape in particular of the extensively studied dielectric loss may strongly be influenced by dipolar cross-correlations. Thus, correlations of \emph{different} molecular dipole moments $\vec\mu_i$: $\langle \vec\mu_i(t)\cdot\vec\mu_j(0)\rangle,\ i\neq j$ may contribute to the overall correlation function in a different way as self correlations $\langle \vec\mu_i(t)\cdot\vec\mu_i(0)\rangle$ do~\cite{pabst2021generic,Boehmer2024b,henot2023orientational,koperwas2022computational}. For this reason, a discussion of the physical origin and the implications of relaxation stretching needs to be refined at the point, where it was previously assumed that self correlations dominate the spectral shape. On this basis, a detailed comparison of dielectric and light scattering spectra recently demonstrated that the broadening in the self correlations strongly favors a high-frequency power law exponent of $\beta=-1/2$, suggesting a generic relaxation stretching of orientational self-correlations in deeply supercooled liquids~\cite{pabst2021generic}. Thus, in the present article we discuss some perspectives and implications of these recent findings and how they relate to common concepts connected to relaxation stretching, like dynamic heterogeneity and widely discussed correlations of the relaxation time distribution with other aspects of supercooled liquid dynamics.  

The manuscript is organized as follows: The introduction covers some general or less recent aspects, including an overview of prior work on the physical origin of relaxation stretching, a discussion on how relaxation stretching can be quantified, a brief glance at the temperature evolution of the spectral shape from the liquid state to the glass transition and finally, an overview of some of the proposed correlations with relaxation stretching. In Section \ref{sec:generic}, we review evidence for a generic relaxation stretching with high-frequency power law behavior $\nu^{-1/2}$. The physical origin for the apparent discrepancy with results from dielectric spectroscopy is addressed in Section \ref{sec:cross-corr}, where we review evidence of how orientational cross-correlations can affect the spectral shape of the structural relaxation. In Section \ref{sec:intra}, we explore how intramolecular dynamics can affect the spectral shape. Finally, we discuss five open questions in Section \ref{sec:perspectives}, which we consider important perspectives for future work.

\subsection{On the physical origin of relaxation stretching}
The physical origin of the asymmetrically broadened structural relaxation peak has been matter of intense debate for decades. While in the following a brief summary of selected experimental and computational results is given, we emphasize that these matters have been discussed in more detail in several excellent reviews, e.g. by Böhmer~\cite{Boehmer1998,Boehmer1998b}, Ediger~\cite{Ediger2000}, Richert~\cite{Richert2002}, Richert et al.~\cite{Richert2011} and Sillescu~\cite{Sillescu1999}. Moreover, we note that the discussion refers to the spectral shape of the structural relaxation or $\alpha$-process, while additional high-frequency relaxation contributions are not considered explicitly.

Regarding the origin of relaxation stretching in deeply supercooled liquids, two limiting cases have been discussed: the heterogeneous and the homogeneous scenario (cf. Refs.~\cite{Boehmer1998,Sillescu1999,Richert2002} and references therein). In the heterogeneous scenario it is assumed that relaxation times of particles are distributed according to some distribution $G(\ln\tau)$, while each particle relaxes exponentially and contributes to the susceptibility spectrum as a Lorentzian, thus being associated with one single relaxation time. The asymmetrically broadened relaxation peak is then obtained as an ensemble average over all particles (see Eqs.~(\ref{equ:glntau}) and (\ref{equ:glntau-t})), thus, in this scenario relaxation stretching is assumed to originate from \textit{dynamic heterogeneity}. On the other side, in the homogeneous scenario the relaxation of each particle is assumed to be intrinsically stretched, while particle relaxation times are not distributed. Of course, also numerous in-between scenarios can be considered that combine a certain degree of intrinsic stretching and some distribution of relaxation times.

Unfortunately, many commonly used experimental techniques are not suited to distinguish between these different scenarios, as they probe the average over all particles in the sample using two-time correlators, or respective linear response equivalents. However, over the years several experimental approaches have been developed that are suited to confirm the existence of dynamic heterogeneity on the particle level. The general idea of early experiments was to selectively probe only a certain dynamic sub-ensemble of particles, e.g. by applying a low-pass filter and probing only the slower-than-average portion of particles. This can be done either by using four-dimensional nuclear magnetic resonance (NMR) experiments, where certain stimulated echo sequences allow for such filtering~\cite{Boehmer1998,Boehmer1998b,Hinze1998,Hinze1998b,Heuer1995}; or by deep photo-bleaching, where fast-relaxing probe molecules dispersed in a supercooled liquid are selectively destroyed and, subsequently, the relaxation of the remaining probes is observed~\cite{Ediger2000,Cicerone1995,Wang1999}. Both experimental approaches found significantly slower relaxation for the obtained sub-ensemble than for the entire system. This observation could only be rationalized in terms of dynamic heterogeneity on the particle level. Similar conclusions could be drawn from dielectric hole burning~\cite{Schiener1996} and solvation dynamics experiments~\cite{Richert1998}. Moreover, it was found that after waiting sufficiently long, the slow subset displayed the average behavior, thus indicating that rate exchange limits the lifetime of dynamic heterogeneity~\cite{Ediger2000,Wang1999,Boehmer1998}. More recently, insightful contributions to the ongoing debate came from fluorescence microscopy experiments~\cite{Paeng2015}. While these experiments have to rely on indirectly probing supercooled dynamics using probe molecules, they offer the unprecedented opportunity to study rotational dynamics of single molecules. Collecting relaxation data of many probe molecules revealed a continuous distribution of relaxation times, thus confirming the established picture of dynamic heterogeneity~\cite{Paeng2015}.

While these experimental findings seem to rule out the purely homogeneous scenario, they are well in line with all in-between scenarios, where some degree of intrinsic broadening instead of pure exponential relaxation exists on the particle level. In this regard, experimental findings are more ambiguous: On the one hand the degree of intrinsic broadening can be estimated indirectly from solvation dynamics~\cite{Wendt2000}, dielectric hole burning~\cite{Schiener1996,Chamberlin1996,Blochowicz2005} as well as NMR experiments~\cite{Boehmer1998}. Here, the results seem to be most compatible with the purely heterogeneous scenario. Noteworthy, this does not necessarily hold for polymers~\cite{Heuer1999,Richert2002} and binary mixtures~\cite{Blochowicz2005}, where the results suggest some degree of intrinsic broadening, possibly related to chain-connectivity effects in the former and local concentration fluctuations in the latter case. On the other hand, single-molecule studies explicitly observe stretched exponential relaxation of probe molecules. However, these results are difficult to interpret, as substantial temporal averaging is required to obtain adequate data quality~\cite{Paeng2015}. Thus, the observed intrinsic broadening of the relaxation could be a manifestation of rate exchange, in the sense that initially fast relaxing probes become slow during the window of observation and vice versa. This conjecture is supported by the observation that the degree of intrinsic broadening is reduced when shorter windows of observation are analyzed. 

More insight is provided with recent advances in computer simulations, which allow to study relaxation of single particles or small clusters \textit{with temporal resolution} using isoconfigurational averaging, i.e. averaging over parallel simulation runs starting with identical particle positions but varying particle velocities~\cite{Berthier2021,DiazVela2020}. It was found that the relaxation of fast particles tend to display some intrinsic broadening, while slow particles relax close to exponential. These results indicate that intrinsic relaxation broadening results from dynamic heterogeneity itself, in the sense that the onset of rate exchange makes single particles experience slightly different heterogeneous environments already during a single relaxation cycle, which results in slightly stretched relaxation \cite{DiazVela2020}.

A very recent study used machine learning methods to accelerate molecular dynamics simulations in order to be able to access times scales on which glassy dynamics could be observed with first-principles accuracy \cite{pabst2024arxiv}. In this way, it was possible to obtain relaxation spectra in excellent agreement with experiments, the relaxation stretching of which was found to be independent of temperature. Interestingly, it was shown that at the same time the dynamic heterogeneity drastically increases on lowering the temperature from being basically absent near the boiling point. This shows that relaxation stretching at the highest temperatures is not due to dynamical heterogeneity and that the relaxation shape is unsensitive to an increasing dynamic heterogeneity, at least up to timescales of several hundreds of nanoseconds.  

Another ongoing debate concerns the spatial nature of dynamic heterogeneity. More specifically the question is as to whether particle mobility is clustered into slow and fast regions associated with some characteristic length scale. While this is suggested for supercooled molecular liquids by some experiments~\cite{Tracht1998,Russell2000,Reinsberg2001}, a large part of the experimental evidence comes from microscopic imaging experiments on colloidal glasses~\cite{Weeks2000,Kegel2000,Mishra2013,Zheng2014}, which could be assumed to resemble molecular glasses in terms of dynamic heterogeneity. Noteworthy, recent advances in computer-simulations have opened the pathway to obtain in-depth information on spatial dynamic heterogeneity in deeply supercooled liquids. This is possible by applying swap Monte-Carlo algorithms~\cite{Ninarello2017} that allow to equilibrate supercooled liquids down to the experimental glass-transition temperature and below by performing unphysical particle swaps. With this advance, the fundamental issue of computer simulations being unable to reach the deeply supercooled temperature range where typical experiments are performed is resolved, allowing to extend the existing evidence from mildly~\cite{Glotzer2000} to deeply supercooled liquids. First pioneering studies following such approaches confirmed the picture of pronounced spatial dynamic heterogeneity, suggesting a coexistence of slow and fast regions and the importance of dynamic facilitation~\cite{Scalliet2022,Guiselin2022}.

\subsection{Quantifying the spectral shape of structural relaxation}
\begin{figure}[h]
\includegraphics[width=0.95\textwidth]{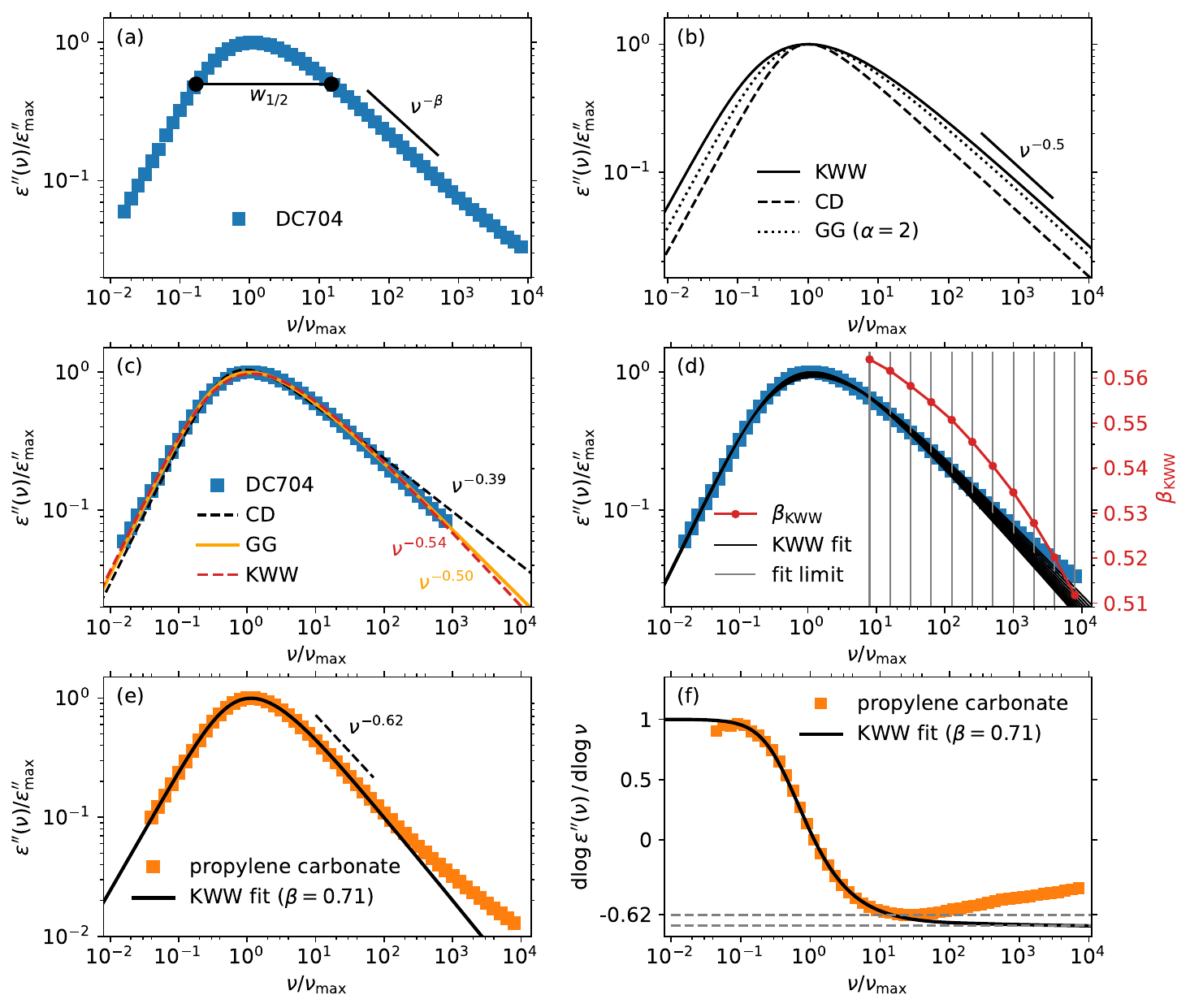}
\caption{\label{fig:model} Quantifying the spectral shape of the structural relaxation. (a) Exemplary dielectric-loss spectrum of the silicone oil DC704 illustrating the definition of the half-width $w_{1/2}$ and the power law exponent $\beta$. (b) Different model functions yield a different peak width despite having the same high-frequency power law (here $\beta=$0.5). (c) Fitting the same data set by using different model functions leads to different $\beta$ values. This is a result of the different inherent spectral shapes of these models. (d) Influence of the fitting range on the resulting stretching parameter in case of the Fourier transform of a KWW function. The grey lines indicate the high-frequency limit of the fitting range, while the red symbols indicate the resulting values of $\beta$ as a function of the high-frequency fitting limit. (e) Comparing the fit of a KWW function with the derivative method for a spectrum of propylene carbonate. The KWW overestimates $\beta$ considerably, due to only reaching its maximum slope beyond the onset of the high-frequency excess wing. (f) Double logarithmic derivative of the data shown in (e) together with the corresponding derivative of the fit function. Applying Eq.~\ref{equ:beta} yields a model-free estimation of $\beta=0.62$.} 
\end{figure}

In order to compare spectral shapes among different supercooled liquids, or to study their temperature dependence, the spectral shape of the structural relaxation has to be quantified using a set of well-defined parameters. The most commonly considered parameter is the high-frequency power-law exponent $\beta$. Occasionally, also the logarithmic half-width $w_{1/2}$ is analyzed. Both are illustrated in \cref{fig:model}(a) for an exemplary spectrum of the silicone oil DC704. Different procedures have been devised to determine these shape parameters from experimental data, including curve fitting procedures and model-free approaches, which all come with certain pros and cons.

\subsubsection{Curve-fitting}
Curve-fitting procedures rely on empirical model functions that have proven to reasonably well describe the spectral shape of supercooled liquids. Besides parameters for the characteristic peak frequency and the overall amplitude, most of the commonly used model functions come with only one single free parameter to control the spectral shape. Consequently, using these models does not allow to vary $w_{1/2}$ and $\beta$ independently, meaning a fit by these models usually represents a compromise of accurately describing the width and the high-frequency behavior of the experimental data. Tuning $\beta$ and $w_{1/2}$ by two independent free parameters is only possible for very few model functions, like, e.g., the one based on a generalized gamma distribution of relaxation times \cite{Blochowicz2003} or on the incomplete gamma function \cite{Kahlau2010a}. 

In \cref{fig:model}(b), commonly used model functions are displayed, including the Fourier transform of the Kohlrausch-William-Watts (KWW) stretched exponential function $\Phi_\text{KWW}\propto \exp{-(t/\tau)^{\beta_\text{KWW}}}$ \cite{Kohlrausch1854}, the Cole-Davidson function (CD) \cite{Davidson1951a} and the relaxation function based on a generalized gamma distribution of relaxation times (GG) \cite{Blochowicz2003} with $\alpha=2$. For all model functions in \cref{fig:model}(b) the high-frequency power law exponent is $\beta=0.5$. Despite that, the shape of the resulting peak varies strongly between the different models. As a consequence of that, the values of $\beta$ obtained by applying curve fitting procedures to experimental data commonly depends on the chosen model. This fact is illustrated in \cref{fig:model}(c), where fitting the loss peak to data produces values of $\beta\in[0.39,0.54]$ depending on the model function. Also fitting in the time instead of the frequency domain, and on a linear instead of a logarithmic scale, can lead to different results due to a different weight of the residuals applied in these cases.

Another aspect to consider is that if the intrinsic peak shape of the model function does not perfectly match the shape of the loss data, then the resulting parameters sensitively depend on the fitting range. This is demonstrated in \cref{fig:model}(d), where the Fourier transform of $\Phi_\text{KWW}(t)$ was applied to the example data set with varying fit ranges resulting in a significant and systematic variation of the resulting exponent $\beta_\text{KWW}$. This aspect is especially important to consider, because high-frequency relaxation contributions like a $\beta$-process or an excess wing represent a "natural" limit to the applicable fitting range. Such a scenario is illustrated in \cref{fig:model}(e) for dielectric loss data of propylene carbonate that feature a distinct excess wing contribution at $\nu/\nu_\mathrm{max}>10^2$. Fitting these data by the Fourier transform of a KWW yields $\beta_\mathrm{KWW}=0.71$, while fitting the high-frequency side of the relaxation peak by a power law gives a much lower $\beta=0.62$. The apparent discrepancy can be explained considering that the KWW fit reaches its limiting $\sim\nu^{-0.71}$ behavior only at $\nu/\nu_\mathrm{max}\sim 10^3$, i.e., at frequencies where the structural relaxation peak can not be disentangled from the excess wing contribution. In our opinion, it is not reasonable to assign $\beta_\mathrm{KWW}=0.71$ despite the experimental data never visibly reaching such a steep power law behavior. 

Thus, while fitting procedures can be valuable tools to analyze relaxation spectra of supercooled liquids over the entire available frequency range, the resulting parameters need to be considered  with great care. An alternative and more straightforward procedure is to extract the high-frequency power law exponent directly from the data without applying any model.

\subsubsection{Derivative analysis}
As an alternative to fitting procedures, the model-free derivative analysis introduced by Nielsen et al.~\cite{Nielsen2009} can be used to obtain more stable results for the high-frequency power law exponent. Here, $\beta$ is defined as the power law tangent at the frequency of steepest slope of the high-frequency side of the relaxation spectrum,
\begin{equation}
    \beta = -\min\left( \frac{\mathrm{d}\,\log\varepsilon^{\prime\prime}}{\mathrm{d}\,\log\nu}\right).
    \label{equ:beta}
\end{equation}
This procedure is illustrated in \cref{fig:model}(f), where the logarithmic derivative spectrum of dielectric data of propylene carbonate from (e) is shown. Eq.~\ref{equ:beta} reveals the steepest slope to be $\beta\approx 0.62$. In accordance to what has been discussed above, the derivative analysis also confirms that the Fourier transform of a KWW function overestimates the high-frequency slope. The derivative of the fit is included in (f), showing that the power law of $\nu^{-0.71}$ is only reached at frequencies considerably beyond the onset of the high-frequency excess wing and, thus, does not correctly reflect the data. The disadvantage of using the derivative approach is that there is no way to quantitatively separate the effects of high-frequency relaxation contributions. Therefore, the best results are obtained at temperatures close to $T_\mathrm{g}$, where the dynamic separation between the structural relaxation process and high-frequency contributions is maximal.

\subsection{Evolution of the spectral shape with temperature}
\begin{figure}[ht]
    \includegraphics[width=0.95\textwidth]{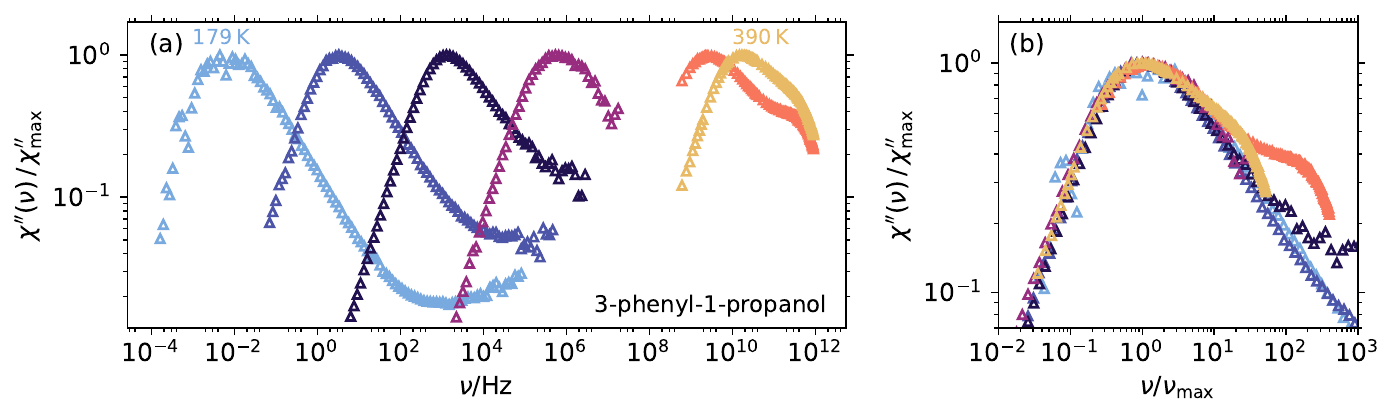}
    \caption{\label{fig:tts}(a) Normalized DLS susceptibility spectra of 3-phenyl-1-propanol from $T_\mathrm{g}\approx 179$ to far above the melting point. (b) Same data but normalized with regard to the respective peak maximum frequencies revealing an approximately temperature-independent shape of the $\alpha$-relaxation.}  
\end{figure}

From slightly below the boiling point down to the glass transition temperature structural relaxation in glass forming liquids slows down continuously covering typically 15 orders of magnitude in relaxation time. During that process the \textit{overall} spectral shape changes dramatically, as different processes show different activation energies. Thus, processes usually separate upon cooling and so the overall composition of the frequency dependent loss drastically changes from all processes being superimposed in the THz regime above and around the melting point to relaxations being spread out over the full dynamic range of 15 decades and more around the glass transition. Whether within this drastic overall change the structural relaxation or $\alpha$-process changes its shape as a function of temperature has been a matter of debate. Generally speaking, this is difficult to tell, as the $\alpha$-process to varying degrees overlaps with other relaxation modes in different temperature regimes. \cref{fig:tts} shows as an example depolarized light scattering data of 3-phenyl-1-propanol in the temperature range 179\,K to 390\,K ($T_\mathrm{g}\approx179\,$K, $T_\mathrm{m}=255\,$K). As the superposition of relaxation peaks reveals in \cref{fig:tts}(b) the $\alpha$-peak shapes only show little changes as function of temperature, a fact that is widely known as frequency-temperature superposition (FTS), or time-temperature superposition (TTS)~\cite{Olsen2001,Niss2018}. Due to the strong overlap with the microscopic dynamics at high temperatures and secondary relaxation processes in the deeply supercooled liquid, however, whether or not FTS is found to hold exactly, or what the deviations precisely are, strongly depends on the approach that is applied to analyze the spectral shape. Thus, in some cases opposite conclusions are drawn in the literature even for the same set of data \cite{Blochowicz2006a,Brodin2007a,Gainaru2009a}. Interestingly, deviations from FTS in the deeply supercooled regime are mostly reported for dielectric data \cite{Ngai2001a, Dixon1990a, Blochowicz2003b}.

\begin{figure}
   \includegraphics[width=0.98\textwidth]{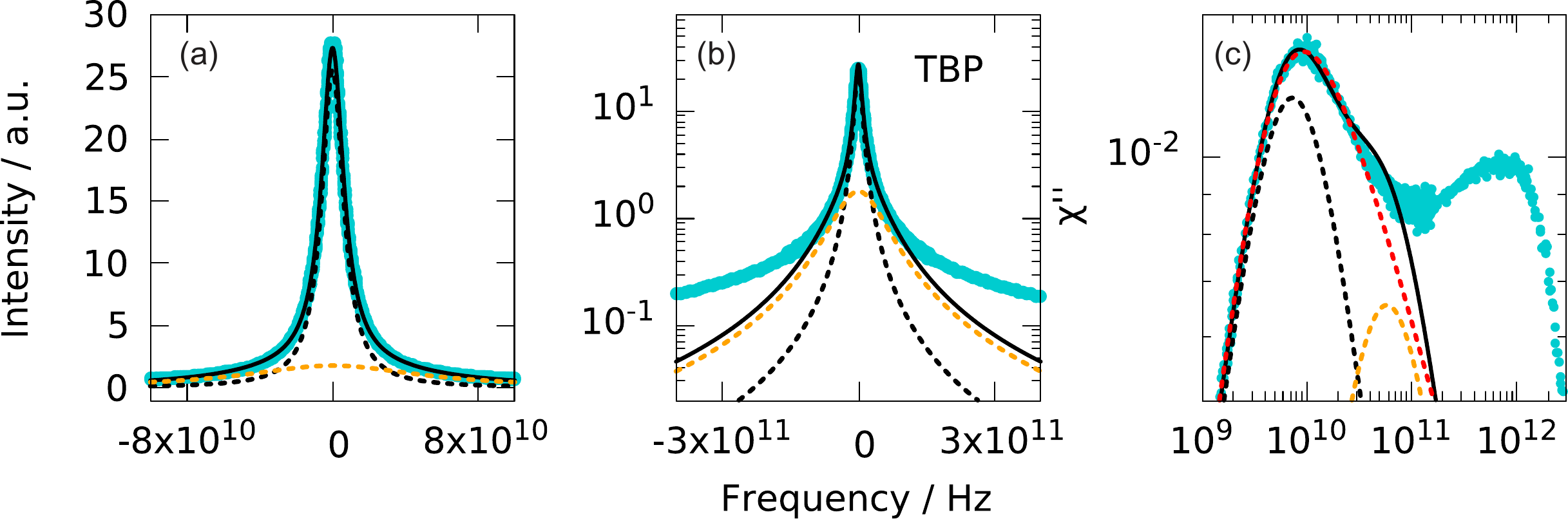}
\caption{\label{fig:tandem} Different representations of a typical depolarized light scattering spectrum of a liquid (tributyl phosphate) above the melting point. Scattering intensity as a function of frequency in linear-linear (a) and semilogarithmic (b) representation. (c) Only the double logarithmic susceptibility representation clearly reveals the broadening of the structural relaxation. Figure adapted from Ref.~\cite{pabst2022intensity}. }
\end{figure}

We also note here that already an approximate validity of FTS clearly challenges the common notion \cite{lunkenheimer2000glassy, Schoenhals1993a} that orientational correlation becomes exponential in the liquid regime due to decreasing spatial dynamic heterogeneity: That the relaxation peak is stretched even in the high temperature regime is not only supported by the observation of wide range of stretching parameters in the temperature range around and above the melting point~\cite{schmidtke2014relaxation}, but also by the fact that for several liquids FTS holds from $T_g$ up to the boiling point~\cite{Petzold2010a, Schmidtke2013a}. One of the reasons for the widespread notion that the main relaxation process becomes exponential above the melting point may be the fact that at high temperatures the broad $\alpha$-relaxation peak merges with the microscopic dynamics. Another reason may be the data representation: Fig.\ \ref{fig:tandem}(a) shows a typical Tandem-Fabry-Perot spectrum of a liquid (tributyl phosphate) above the melting temperature. When the scattering intensity $I(\nu)$ is considered in linear-linear representation a Lorentzian function plus a broad background, the latter thought to be unrelated to rotational motions,  appears to describe the central line reasonably well, suggesting an exponential $\alpha$-relaxation. Only the susceptibility representation in \cref{fig:tandem}(c) clearly reveals the considerable relaxation stretching, which can be described by a CD with high-frequency power law exponent $\beta\approx 0.5$ (red dashed line).  

\subsection{Proposed correlations and scaling of the spectral shape}
Numerous attempts have been made to identify universal characteristics of the spectral shape of the structural relaxation shared by different deeply supercooled liquids, in particular because a generally accepted theory of the glass transition is still lacking. Thus, it was attempted to relate the various values of the stretching exponent $\beta$ to several other characteristics of supercooled liquids. In the following only a few examples are given.

Böhmer et al., e.g., reported a correlation of $\beta$ with the fragility index $m = \left.\mathrm{d}\log\tau/\mathrm{d} (T_g/T)\right|_{T_g}$, a measure for the non-Arrhenius temperature dependence of relaxation times \cite{Boehmer1993a}. The analysis also included polymers and network glasses, while the evidence for supercooled molecular liquids alone is rather weak and thus, has been discussed critically \cite{Dyre2007, Gupta2008}. 

In the coupling model proposed by Ngai et al.\ (see Ref.\ \cite{Ngai2023} and references therein), $\beta$ is assumed to be related to the dynamic separation of the $\alpha$- and the Johari-Goldstein secondary relaxation process. Here, the latter was argued to be associated with the primitive relaxation of the system, i.e., a precursor to the $\alpha$-relaxation dynamics. Mostly, this model was discussed in the context of dielectric data so far. How to include recent results from depolarized light scattering, nuclear magnetic resonance or shear mechanical spectroscopy is not directly clear. 

A universal relation of the $\alpha$-peak width with the high frequency excess wing was suggested by a scaling procedure proposed by Nagel and coworkers~\cite{Dixon1990a}. The excess wing feature appears as a deviation from the simple power law behavior $\nu^{-\beta}$ on the high frequency side of the dielectric $\alpha$ peak. Although the scaling proved to work to a certain extent, formal objections were raised \cite{Kudlik1995a,Leheny1996a,Kudlik1996a} and modifications were suggested to circumvent these \cite{Dendzik1997a,Paluch1998a}, but also alternative approaches were proposed by scaling a certain set of fit parameters instead of the data \cite{Blochowicz2003,Blochowicz2006a} or by scaling the data onto a common envelope \cite{Gainaru2019}. Later on, however, physical aging experiments~\cite{Schneider2000a} and data on binary mixtures~\cite{Blochowicz2004a} indicated that the excess wing could also be considered as a secondary relaxation that is submerged with the $\alpha$ process, and, thus, should be treated separately. 

Moreover, as $\beta$ oftentimes is considered as a measure for the degree of dynamic heterogeneity, its relation to the length scale of dynamic heterogeneity has been studied using NMR techniques by Qui et al.\ \cite{Qiu2003}. Although the somewhat expected correlation of smaller values of $\beta$ being associated with a larger dynamic heterogeneity length scale was observed, the study considered only four substances with quite different molecular structures and even included a polymer, thus no definite conclusions can be drawn.

Nielsen et al.~\cite{Nielsen2009} evaluated possible correlations between $\beta$ and temperature, quality of the power law fit, validity of TTS, $w_{1/2}$, degree of non-Arrhenius temperature dependence of relaxation times and dielectric relaxation strength $\Delta\varepsilon$. Their analysis of dielectric data of a large number of organic glass-formers revealed a special status for liquids with $\beta=1/2$, which, e.g., tend to strictly obey TTS. 

Paluch et al.\ \cite{Paluch2016} reported a correlation between $\beta$ and $\Delta\varepsilon$ for a broad variety of supercooled molecular liquids. Here, larger $\beta$ values are associated with a larger relaxation strength, meaning the relaxation spectra of highly polar liquids tend to be less stretched than those of less polar ones. Initially, this observation was rationalized by arguing that strong dipolar interactions increase the harmonicity of the intermolecular potentials, thus affecting dipolar self-correlations \cite{Paluch2016}. More recent results, however, indicate that dipolar cross correlations play a major role as dipolar interactions become increasingly important~\cite{Boehmer2024b}. This will be discussed in more detail further below.

Thus, despite several attempts to find universal features in the temperature evolution of the spectral shape of the $\alpha$ process, the results so far have not been clear enough to lead to a generally accepted picture of the temperature evolution of the spectral shape of the reorientation dynamics in supercooled molecular liquids and possible correlations with various properties of supercooled liquids keep being controversially debated. 

\section{The generic spectral shape of structural relaxation}
\label{sec:generic}
\begin{figure}[h]
   \includegraphics[width=0.48\textwidth]{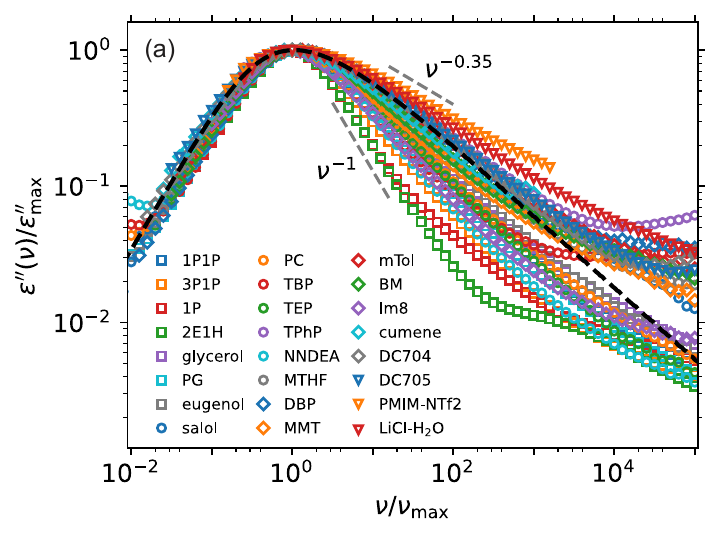}
   \includegraphics[width=0.48\textwidth]{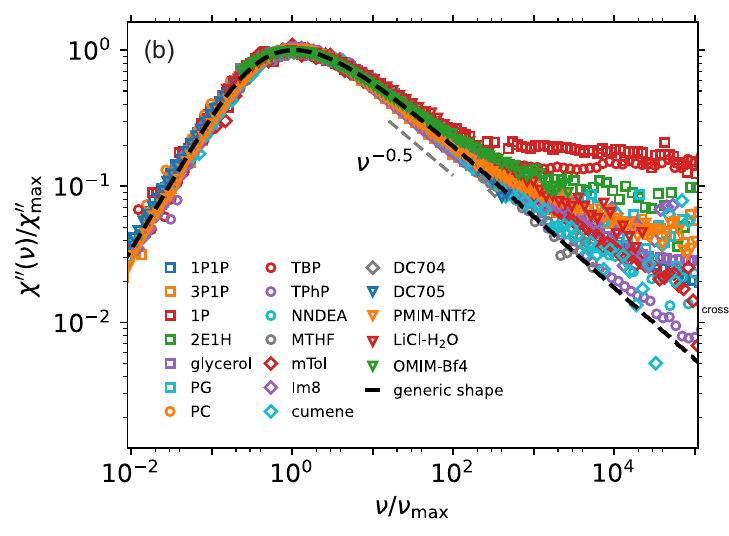}
   \caption{  \label{fig:master} Comparing the spectral shapes of different supercooled liquids. (a) Dielectric loss $\edp$ and (b) DDLS spectra $\cdp$. All spectra are normalized with regard to peak-maximum frequency and amplitude. The shape of the dielectric loss varies strongly among the different systems, with high-frequency power-law exponents ranging from 0.35 to 1.0. On the other hand, all light-scattering data approximately collapse onto a single master curve, which is well-described by the spectral shape based on the GG distribution of relaxation times with parameters $\alpha=2.0$ and $\beta=0.5$ (dashed black line). Abbreviations and reference to original data: 1-phenyl-1-propanol~(1P1P)~\cite{Boehmer2019}, 3-phenyl-1-propanol~(3P1P)~\cite{Boehmer2019}, 1-propanol~(1P)~\cite{Gabriel2017}, 2-ethyl-1-hexanol~(2E1H)~\cite{Boehmer2023}, propylene glycol~(PG)~\cite{bohmer2022glassy}, propylene carbonate~(PC)~\cite{pabst2021generic}, tributyl phosphate~(TBP)~\cite{Pabst2020}, triethyl phosphate~(TEP), triphenyl phosphite~(TPhP)~\cite{pabst2021generic}, N,N-diethylacetamide~(NNDEA), 2-methyltetrahydrofuran~(MTHF)~\cite{pabst2021generic}, dibuthyl phthalate~(DBP), methyl-m-toluate~(MMT), m-toluidine~(mTol) (dielectric data from~\cite{Mandanici2005}), butyl methacrylate~(BM), 1-octylimidazole~(Im8), 17\% LiCl-water solution(LiCl-H$_2$)~\cite{pabst2021generic}, salol~\cite{Casalini2003}, glycerol~\cite{Gabriel2020}.} 
\end{figure}

For several decades, dielectric spectroscopy has been the predominant experimental technique to study the spectral shape of deeply supercooled organic liquids. One of the main reasons for this is the superior data quality, as, e.g., reliable values of $\beta$ can only be extracted with good accuracy on a logarithmic scale, requiring excellent signal-to-noise ratio. As a consequence, several meta-analyses comparing the spectral shape of different supercooled organic liquids considered mainly dielectric-loss data. There it was found that $\beta$ (and also $w_{1/2}$) varies significantly among different supercooled liquids~\cite{Nielsen2009,Paluch2016,Boehmer1993}. This variety of dielectric-loss spectral shapes is illustrated in \cref{fig:master}a), comparing data of a broad variety of supercooled organic liquids normalized to the respective peak frequencies and amplitudes. Among the presented selection, $\beta$ varies from almost 1 for some monohydroxy alcohols to 0.35 for ionic liquids.

Contrary to these observations from dielectric spectroscopy, in recent years growing evidence supports the conjecture of a generic spectral shape of the $\alpha$-process \cite{pabst2021generic}. These observations have been enabled by advances in other experimental techniques, e.g. dynamic light-scattering and rheology, reaching experimental resolutions comparable to the previously superior dielectric experiments. Fig.\ \ref{fig:master}b) presents depolarized light-scattering data of a broad variety of substances normalized to the peak maximum frequency and amplitude. Evidently, all these data follow to good approximation the dashed black line, featuring a high-frequency power law $\nu^{-1/2}$. Major deviations are found only at $\nu/\nu_\mathrm{max}>10^2$, where secondary relaxation processes start to contribute, the appearance of which is individual for each substance. Fig.\ \ref{fig:master}b) contains data of monohydroxy alcohols, polyhydric alcohols, polar and apolar van der Waals liquids and ionic systems, thus it covers the full spectrum of organic low-molecular-weight glass formers. Recently, the same exponent $\beta\approx 0.5$ was observed for the inorganic glass-former GeO$_2$~\cite{Sidebottom2023} and mixtures with Na$_2$O~\cite{Sidebottom2024} close to $T_\mathrm{g}$ using polarized dynamic light-scattering, thus suggesting that the observed generic behavior might also apply for inorganic substances.

Similar conclusions were drawn from rheology experiments: Bierwirth et al.~\cite{Bierwirth2017} found that the fluidity spectra of several supercooled molecular liquids have very similar shapes, well-described by the random barrier model by Dyre et al.~\cite{Dyre2000}. Later, this observation was extended to an even broader range of glassy materials by Gainaru et al.~\cite{Gainaru2019}, including polymeric, inorganic and metallic melts. The fluidity is the mechanical equivalent of the electric conductivity, thus these results can not directly be compared to the relaxation spectra from dielectric spectroscopy or light-scattering discussed above. Although fluidity data could be transformed into the compliance representation to allow for a direct comparison in principle, experimental difficulties often prevent this procedure.

\begin{figure}[h]
    \includegraphics[width=0.4\textwidth]{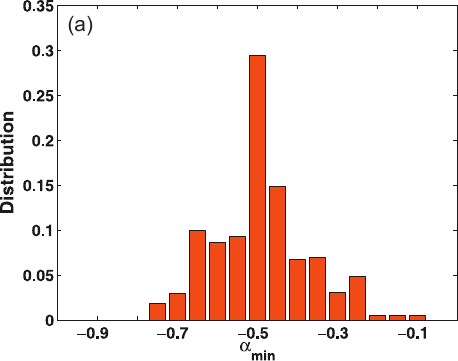}\hspace{1cm}
   \includegraphics[width=0.4\textwidth]{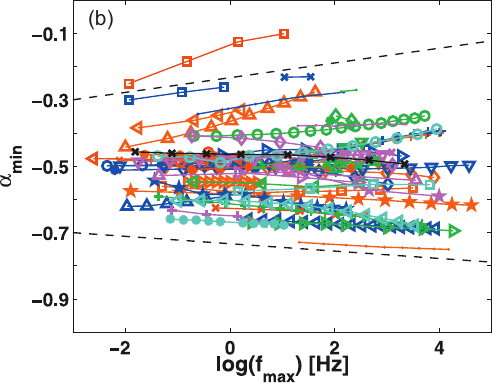}
   \caption{\label{fig:nielsen}Figure from Nielsen et al.~\cite{Nielsen2009}, reporting the high-frequency power law exponent as a function of peak-maximum frequency extracted from dielectric-loss data of various organic glass-formers. High-frequency power law exponents were extracted using the derivative technique (note that $\alpha_\mathrm{min}\equiv \beta$). Figure reproduced from Ref.~\cite{Nielsen2009} with the permission of AIP Publishing.} 
\end{figure}

In the light of all the evidences supporting the conjecture of a generic spectral shape with $\beta\approx 0.5$, the question arises why in some cases considerably different spectral shapes are observed in dielectric experiments, especially for strongly polar liquids, as shown in Fig.~\ref{fig:master}(a). In fact, Nielsen et al.~\cite{Nielsen2009}  studied the distribution of $\beta$-values from dielectric spectroscopy among many supercooled organic liquids. Most notably, they found that the median of the distribution corresponds to $\beta\approx 0.5$, see \cref{fig:nielsen}a. As shown in \cref{fig:nielsen}, substances with $\beta$ -values larger or smaller than $1/2$ were found to converge towards $\beta=0.5$ when approaching low temperatures close to $T_\mathrm{g}$, as suggested already earlier by Olsen et al~\cite{Olsen2001}. While these results indicate that the $\nu^{-1/2}$ high-frequency behavior is of special significance also for dielectric-loss data, they do not provide an explanation for strong deviations from $\nu^{-1/2}$ observed in some instances. Also, they explicitly excluded substances with high dielectric loss from their analysis, for which the deviations are even more pronounced. Recently, however, we proposed that contributions from dipolar cross-correlations seem to be the origin of these deviations~\cite{pabst2021generic,Boehmer2024b}. Evidence for this conjecture will be discussed in the following section.

\section{The role of cross-correlations}
\label{sec:cross-corr}

We start by pointing out how dipolar cross-correlations contribute to the dielectric loss, which is best illustrated by considering the time-autocorrelation function of electric polarization fluctuations of the sample, 
\begin{equation}
    \label{equ:polcorr}
    C_{\bm{P}}(t) = \langle \bm{P}(0)\cdot\bm{P}(t)\rangle.
\end{equation}
Via fluctuation-dissipation and linear-response relations, $C_{\bm{P}}(t)$ is related to the dielectric loss $\edp(\omega)$ via inverse Laplace transform,
\begin{equation}
    \edp(\omega) = \frac{\de}{\omega}\int\limits_0^\infty C_{\bm{P}}(t) \cos\omega t\,\dd t.
\end{equation}
By relating the macroscopic polarization fluctuations to the microscopic fluctuations of permanent dipole moments, one obtains
\begin{equation}
    \label{equ:selfcross}
    C_{\bm{P}}(t) = \frac{1}{V^2}\Bigl\langle\sum\limits_i \bmu_i(0)\cdot\sum\limits_j \bmu_j(t)\Bigr\rangle = \frac{1}{V^2}  \sum\limits_i \bigl\langle\bmu_i(0)\cdot\bmu_i(t)\bigr\rangle  + \frac{1}{V^2}\sum\limits_{i\neq j} \bigl\langle\bmu_i(0)\cdot\bmu_j(t)\bigr\rangle,
\end{equation}
where $\bmu_i(t)$ is the molecular dipole moment vector of molecule $i$ at time $t$. In the last part of \cref{equ:selfcross} one can distinguish two contributions: dipolar self-correlations, \ie orientational time-autocorrelations of the single molecular dipoles, and dipolar cross-correlations, quantifying the orientational correlation of dipoles with respect to surrounding dipoles. 

It is well established that dipolar cross-correlations can considerably affect the \textit{static} dielectric constant of certain types of supercooled liquids, as quantified by the Kirkwood correlation factor
\begin{equation}
    \label{equ:gk}
    \gk = 1 + \frac{1}{\mu^2}\Bigl\langle\bmu_i\cdot\sum\limits_{i\neq j}\bmu_j\Bigr\rangle = \frac{9k_\mathrm{B}\ez M T}{\rho N_\mathrm{A}\mu^2}\frac{(\es-\ei)(2\es+\ei)}{\es(\ei+2)^2}.
\end{equation}
Here, $\es=\lim_{\nu\to 0}\ep(\nu)$ and $\ei=\lim_{\nu\to \infty}\ep(\nu)$ are the zero and high frequency limits of the dielectric permittivity, $T$ is temperature, $M$ molar mass, $\rho$ mass density and $\mu$ the gas-phase molecular dipole moment~\cite{Kirkwood1939a,Froehlich1958a}. As evident from the second part of \cref{equ:gk}, $\gk>1$ indicates positive orientational cross-correlations due to preferred parallel alignment of adjacent dipoles, while $\gk<1$ indicates the opposite. $\gk=1$ can indicate the absence of any dipolar cross-correlations, however, would also be observed if preferred parallel and anti-parallel alignments coexist.

$\gk>1$ can be observed for different supercooled liquids, the most famous examples probably being monohydroxy alcohols (MAs). In these substances, chain-like supra-structures are formed via hydrogen bonding, resulting in strong orientational correlations between adjacent dipole moments.\cite{Boehmer2014} In MAs with certain molecular architectures, also $\gk<1$ is observed, indicating the prevalence of ring-like supra-structures~\cite{Dannhauser1968a,Dannhauser1968b,Dannhauser1968c,Singh2012}.

While static cross-correlations can easily be identified by calculating $\gk$, determining their dynamic signature is less straight-forward. While the general necessity of considering the dynamic signature of cross-correlations has been discussed early~\cite{Keyes1972,Kivelson1975,Madden1984}, this matter has been mostly ignored in experimental studies and has only recently attracted more attention. In the following sections we argue that dipolar cross-correlations can superimpose the self-correlations as an additional slow and narrow process to the dielectric loss. The spectral shape of the resulting superimposed peak strongly depends on the strength of these cross-correlations. Finally, we discuss how these experimental findings are supported by recent computational work.

\subsection{Identifying cross-correlation contributions to relaxation spectra}
\begin{figure}[h]
   \includegraphics[width=0.95\textwidth]{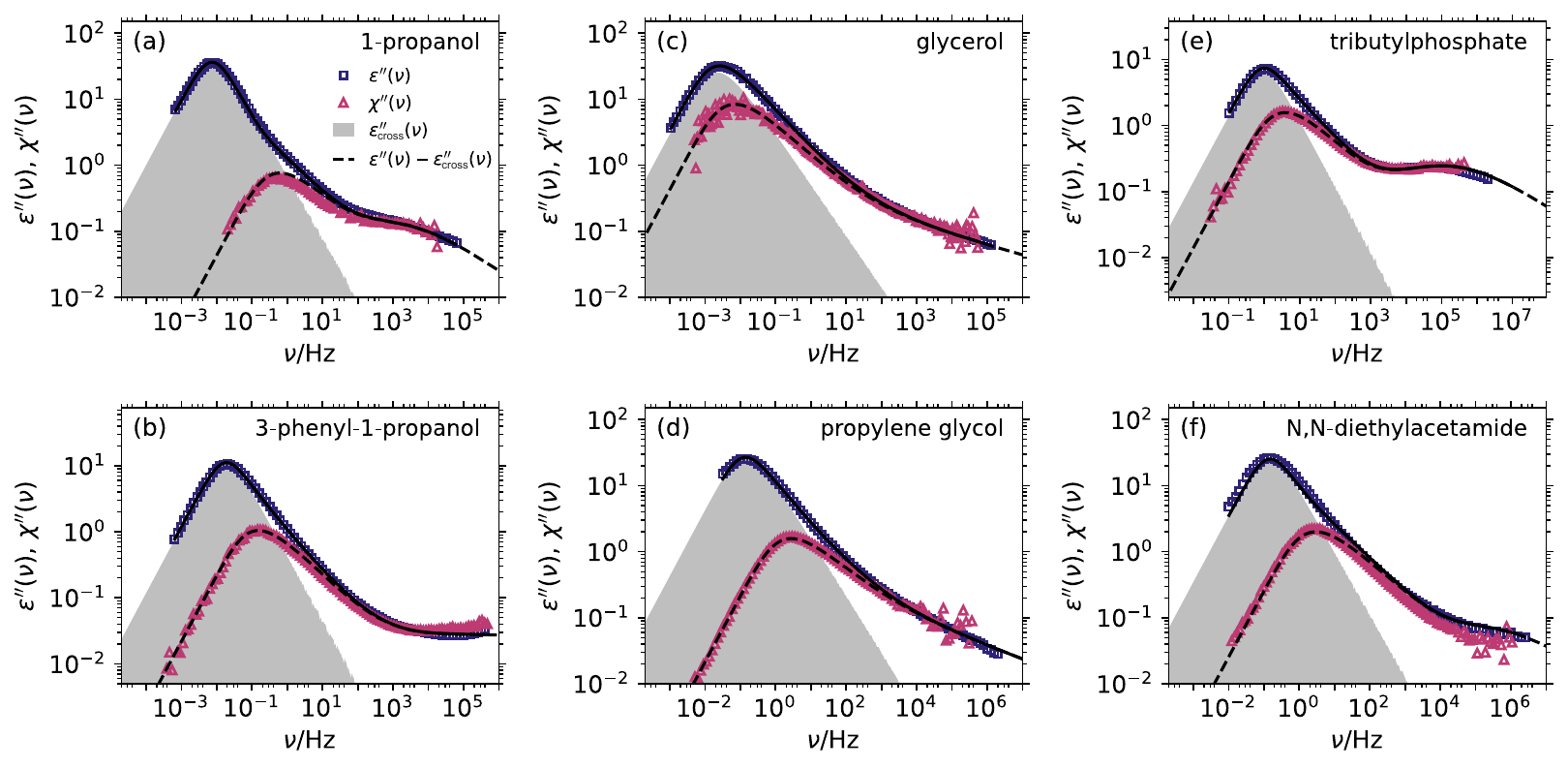}
   \caption{Identifying contributions of dipolar cross-correlations to the dielectric loss (blue symbols) through comparison to light-scattering data (purple symbols). The substances in panels (a) and (b) are monohydroxy alcohols, in panels (c) and (d) are polyhydric alcohols and in panels (e) and (f) are polar van-der-Vaals liquids. The gray shaded areas illustrate the slow cross-correlation contributions and the dashed black lines represent the curve obtained by subtracting the cross-correlation contribution from the dielectric loss.} 
  \label{fig:bds_super}
\end{figure}

In typical MAs with $\gk>1$, like e.g. 1-propanol, the dynamic signature of dipolar cross-correlations has been studied extensively~\cite{Boehmer2014,Hansen1997,Sillren2014,Gainaru2010}. In the dielectric-loss spectrum they appear as the so-called \textit{Debye process}, a narrow peak at significantly lower frequencies compared to self-correlations. The Debye process can be understood as the reorientation of hydrogen-bonded supra-structures. The latter is currently believed to proceed via transient detachments and attachments from/to the supra-structures \cite{Gainaru2010}.

More recently, the empirical observation was made that the self-correlation contribution to the dielectric-loss spectrum can be identified by terms of depolarized dynamic light scattering experiments~\cite{Gabriel2017,Gabriel2018a,Boehmer2019,Boehmer2023}. The procedure is illustrated for two MAs, 1-propanol~\cite{Gabriel2017} and 3-phenyl-1-propanol~\cite{Boehmer2019}, in the left side panels (a) and (b) of \cref{fig:bds_super}. Here, the dielectric loss $\edp(\nu)$ is shown as blue and the light-scattering susceptibility spectrum $\cdp(\nu)$ as purple symbols. While the two spectra coincide at high frequencies, large deviations due to dipolar cross-correlations are observed at low frequencies. Once this Debye process (gray shaded area) is subtracted from the fit to $\edp(\nu)$ (black solid line), the black dashed line is obtain, which almost perfectly coincides with $\cdp(\nu)$. 

MAs are not the only substances with relevant contribution of slow dipolar cross-correlation to the dielectric loss, though. Similar scenarios as known for MAs are observed for the polyhydric alcohols glycerol (c) and propylene glycol (d)~\cite{bohmer2022glassy}, but also for some non-hydrogen bonding polar supercooled liquids like tributyl phosphate~\cite{Pabst2020} (e) and N,N-diethylacetamide (f). Again, for these substances $\edp(\nu)$ is well-described as a superposition of a fit to $\cdp(\nu)$ and a slow Debye-like process. These observations suggests that, besides chain-like hydrogen bonding, also the hydrogen bonded networks in polyhydric alcohols and pure dipole-dipole interactions in polar non-hydrogen bonding substances can induce notable cross-correlation contributions.

In fact, a recent theory of dielectric relaxation by Déjardin et al.\ suggests that $\gk>1$ alone is a sufficient condition to observe a slow cross-correlation contribution to the dielectric loss~\cite{Dejardin2019}. The theory determines the static dielectric permittivity of an isotropic polar fluid from the Langevin equations describing the dynamics of particles~\cite{Dejardin2022a}. Although the theory does not consider any complex intermolecular interactions like hydrogen-bonding, it predicts that $\gk>1$ can be observed for polar liquids simply due to electrostatic interactions of permanent dipoles as well as induction and dispersion forces \cite{Dejardin2022b}. At the same time, earlier versions of the theory considered the dynamic susceptibility, where for $\gk>1$ an additional slow cross-correlation process is observed, similar to what is is found for experimental data~\cite{Dejardin2019}.

The identification of a slow and narrow dipolar cross-correlation contribution to the dielectric-loss spectrum suggests a qualitative explanation for the apparent contradicting results regarding the spectral shape compared to other techniques: While the self-correlations seem to to have a quite generic shape with $\beta\approx 0.5$, dielectric loss peaks representing a superposition of self- and cross-correlations usually are more narrow with $\beta > 0.5$. A corresponding quantitative relation is discussed in one of the subsequent sections.

\begin{figure}[h]
   \includegraphics[width=0.48\textwidth]{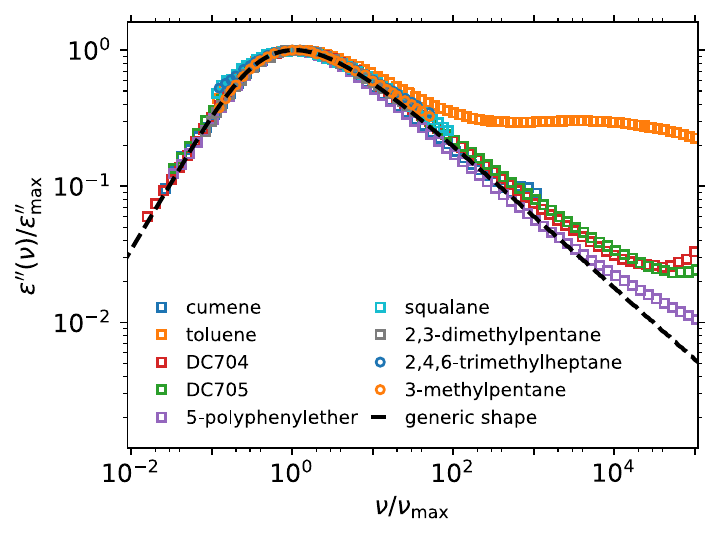}
        \caption{Dielectric-loss data of various low-polarity supercooled liquids. The spectral shapes are very similar and are found to follow a $\nu^{-1/2}$ high-frequency behavior and to correspond to the black dashed line found to describe the light-scattering data in \cref{fig:master}. Data for the branched alkanes, 2,3-dimethylpentane, 2,4,6-trimethylheptane and 3-methylpentane, were obtained from Ref.~\cite{Shahriari2004} and toluene data are from Ref.~\cite{Kudlik1999}.} 
    \label{fig:lowPolar}
\end{figure}

Following these notions, it is worth to take a look at low-polarity supercooled liquids, for which dipole-dipole interactions should play a minor role. Thus, one could expect that in these substances only dipolar self-correlations contribute to the dielectric loss. In line with this notion, \cref{fig:lowPolar} reveals the dielectric loss of a selection of low-polarity substances to follow a $\nu^{-1/2}$ high-frequency behavior. The figure includes aromatic compounds, branched alkanes and silicone oils. 

The evidence based on the comparison of dielectric-loss spectra and dynamic light scattering data are complemented and supported by findings from several other experimental techniques. In particular, $^2$H nuclear magnetic resonance (NMR) plays a crucial role, because this technique unambiguously probes an orientational self-correlation function. Therefore, the interpretation of the corresponding results does not rely on any assumptions, like, e.g., the empirical observation that DDLS is mostly insensitive to orientational cross-correlations. Analyses of NMR data obtained for a wide variety of different supercooled liquids largely support the findings from DDLS, i.e., spectral shapes of the structural relaxation from NMR and DDLS agree well, while in many instances deviations from the dielectric spectral shape are observed~\cite{Koerber2020a,Becher2021,Becher2022}. However, NMR observes a broader variety of high-frequency power-law exponents compared to DDLS. It is difficult to judge, however, whether some of these discrepancies might result from the fitting procedures applied to NMR data. This is supported by the fact that quite different results were obtained using different fit models like KWW and CD~\cite{Koerber2020a}, as discussed above .

Cross-correlation contributions were also identified in comprehensive analyses of shear rheological data. In the case of MAs the shear modulus is dominated by the $\alpha$-relaxation, which is observed at similar frequencies as in DDLS. In addition, a low-frequency contribution is commonly identified that resembles chain-connectivity effects in polymers~\cite{Gainaru2014,Mikkelsen2023} and, thus, is thought to reflects cross-correlations between molecules resulting from hydrogen-bonding. A similar contribution was also identified for glycerol in the complex viscosity representation by Arrese-Igor et al.~\cite{ArreseIgor2020b}. It was found to coincide with the peak-maximum frequency of the dielectric loss spectrum, which supports the conjecture that the latter represents a cross-correlation contribution. Similar conclusions regarding the coexistence of self-correlation and collective modes in shear rheological spectra of glycerol were drawn by Gabriel et al.~\cite{Gabriel2020} for the shear compliance~\cite{Gabriel2020}. In fact, it is not surprising that hydrogen-bonded structures considerably contribute to shear rheological quantities considering that hydrogen bonds have to be broken in order to initiate flow~\cite{Patil2023}. There is no consensus, however, as to whether the same behavior is to be expected for dipole-dipole interactions-induced cross-correlations. For tributyl phosphate (see \cref{fig:bds_super}), Moch et al.~\cite{Moch2022} identified only a single relaxation contribution to the real part of the shear compliance, the average relaxation time of which roughly coincides with the one extracted from the dielectric loss. This, among other observation, lead these authors to the conclusion that the collective relaxation mode instead of the self-correlations are associated with the glass transition. Arresse-Igor et al.~\cite{Arrese-Igor2023} came to a different conclusion, as they clearly identified a self- as well as a cross-correlation contribution to the imaginary part of the viscosity and the derivative of the real part of the shear compliance. The latter can be thought of as an approximation of the imaginary part of the shear compliance. These findings support the light-scattering results discussed above and clearly show that cross-correlational modes contribute differently to the different shear rheological quantities. 

Similar observations were recently reported by Paluch et al., who proposed an approach for identifying dipolar cross-correlations via the dielectric modulus $\hat{M}(\omega)$~\cite{Paluch2023}, i.e., the dielectric response upon applying a constant charge instead of a constant electric field. Because $\hat{M}(\omega)=\hat{\varepsilon}(\omega)^{-1}$, relaxation processes contribute to the dielectric modulus with a different weight compared to the dielectric permittivity. Processes with a rather large $\Delta\varepsilon$, like the dipolar cross-correlation effects in certain supercooled liquids, contribute to $\hat{M}(\omega)$ with a rather small $\Delta M$. As a consequence, the dielectric modulus in many cases is dominated by dipolar self-correlations, as shown by Paluch et al. through comparison with DDLS data~\cite{Paluch2023}. 

\subsection{Suppression of cross-correlations}
\begin{figure}[h]
   \includegraphics[width=0.48\textwidth]{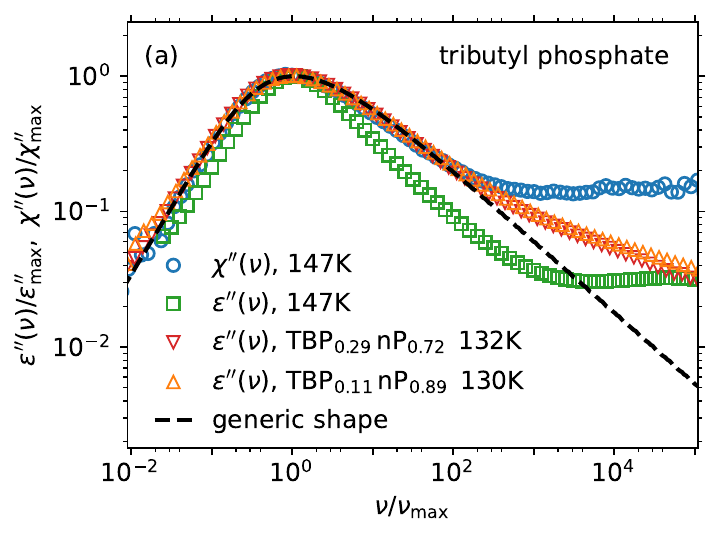}
   \includegraphics[width=0.48\textwidth]{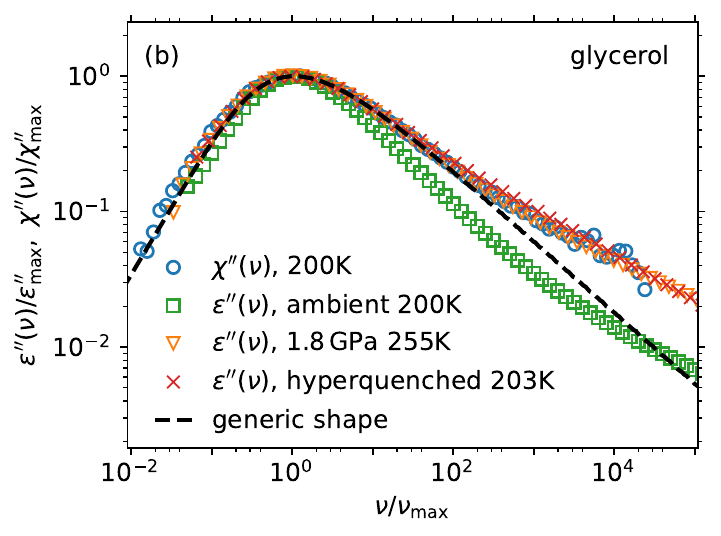}
   \caption{Verifying whether suppressing dipolar cross-correlations eliminates discrepancies between the dielectric-loss and light-scattering data. Panel (a) is reproduced from Ref.~\cite{Pabst2020} and shows dielectric-loss (green symbols) and light-scattering data (blue symbols) of pure TBP, as well dielectric-loss data from two mixtures with n-pentane with different concentrations indicated in the legend (red and orange symbols). Dilution suppresses dipolar cross-correlations and recovers the generic spectral shape (black dashed line). Similar data are shown in panel (b) for glycerol, where hydrogen bonding is suppressed either by applying 1.8\,GPa hydrostatic pressure (orange symbols)~\cite{HenselBielowka2004}, or by hyperquenching from room temperature with $80,000\,$K/s (red symbols)~\cite{Gainaru2020}. Both procedures recover the generic spectral shape.} 
  \label{fig:suppr}
\end{figure}

Following the conjecture that dipolar cross-correlation contributions control the shape of the dielectric loss, the spectral shape of self-correlations should be recovered for the dielectric loss once cross-correlations are suppressed using a suitable experimental procedure. \cref{fig:suppr} presents three examples of such procedures for two different supercooled liquids. Panel (a) reproduces data of tributyl phosphate (TBP) from Ref.~\cite{Pabst2020} for which a distinct slow cross-correlation contribution to the dielectric loss was identified in \cref{fig:bds_super}(e). Attempting to suppress these cross-correlations, Pabst et al.~\cite{Pabst2020} studied mixtures of TBP with the non-polar solvent n-pentane. This procedure is thought to increase the average distance between TBP molecules, thus weakening dipole-dipole interactions. For a direct comparison, all data in \cref{fig:suppr} are normalized to their respective peak-maximum frequencies and amplitudes. A significantly more narrow spectral shape is observed for the dielectric loss (green symbols) compared to the light-scattering spectrum (blue symbols). Sufficient dilution by n-pentane, however, leads to a broadening of the dielectric-loss spectrum, such that it coincides with the shape of the light-scattering spectrum up to frequencies where secondary relaxations dominate the spectral shape. The broadening was confirmed to saturate at high n-pentane concentrations, thus, it can be excluded that it is related to enhanced dynamic heterogeneity in the binary mixture compared to pure TBP.

Panel (b) shows similar results for glycerol, but based on different procedures to suppress cross-correlations, which in glycerol most likely reflect the hydrogen-bonded network. One way to suppress hydrogen-bonding is to apply large hydrostatic pressures~\cite{Naoki1991,Cook1992}, while a second possibility is to hyperquench glycerol from room-temperature directly into the deeply supercooled state. The latter essentially freezes the high-temperature structure of the liquid, as the system is given insufficient time for structural equilibration. Since hydrogen bonding is less pronounced at high temperatures, the state of glycerol obtained by hyperquenching is expected to feature a less pronounced hydrogen-bonded network. In \cref{fig:suppr}(b) dielectric-loss data of pressurized glycerol at 1.8\,GPa from Hensel-Bielowka et al.~\cite{HenselBielowka2004}, as well as hyper-quenched glycerol cooled with $80,000\,$K/s from Gainaru et al.~\cite{Gainaru2020} are compared to data obtained for pure glycerol. Similar as found for TBP, both procedures of suppressing hydrogen bonding lead to an almost perfect collapse of the dielectric loss on the light-scattering data of pure glycerol. In accordance, pressurized and hyperquenched glycerol displays a significantly reduced dielectric relaxation strength and reduced Kirkwood correlation factor compared to ambient conditions~\cite{HenselBielowka2004,Gainaru2020}. Therefore, in line with the fact that the dynamic cross-correlation contributions seems to be suppressed, also static dipolar cross-correlations are reduced.

\subsection{Quantitative relation between high-frequency power law exponent and cross-correlations}

\begin{figure}[h]
   \includegraphics[width=0.48\textwidth]{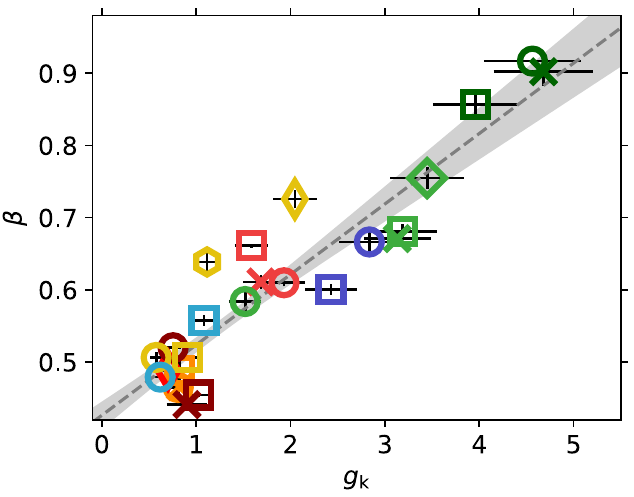}
    \caption{High-frequency power law exponent $\beta$ of the dielectric loss as function of the Kirkwood correlation factor for 25 supercooled liquids. A positive correlation between both quantities is observed (grey dashed line). Symbol colors indicate sub-categories of substances: Orange: phthalates, red: phosphates, dark red: aromatic van-der-Waals liquids, yellow: non-aromatic van-der-Waals liquids, blue: polyhydric alcohols, light blue: phenols, light green: aromatic monohydroxy alcohols, dark green: non-aromatic monohydroxy alcohols. See Ref.~\cite{Boehmer2024b} for more details.} 
    \label{fig:corr_beta_gk}
\end{figure}

While \cref{fig:bds_super} confirms the contribution of dipolar cross-correlations to the dielectric loss spectra of selected supercooled liquids, a quantitative relation between the spectral shape and the strength of dipolar cross-correlations has recently been established for a large set of supercooled liquids~\cite{Boehmer2024b}. \cref{fig:corr_beta_gk} reports the main result, i.e., a positive correlation between the strength of dipolar cross-correlations quantified by the Kirkwood correlation factor $\gk$ and the high-frequency power law exponent $\beta$ determined using the derivative analysis (see \cref{equ:beta}). This result confirms that supercooled liquids with strong dipolar cross-correlations tend to have steeper high-frequency power laws. For supercooled liquids without any relevant cross-correlations ($\gk\approx 1$), however, $\beta\approx0.5$ is found, in accordance to the spectral shape observed in light-scattering experiments. The observed relation between $\gk$ and $\beta$ holds for a variety of different substance classes, which interact via different mechanisms like hydrogen bonding, as well as dipole-dipole and possibly also $\pi-\pi$ interactions. In all cases, the degree of static cross-correlations, quantified by $\gk$, predicts how the dynamic signatures of these cross-correlation affects the spectral shape of the dielectric loss. 

\subsection{Computational evidence}
Computer simulations can serve as valuable tools to unambiguously disentangle dipolar self- and cross-correlations, due to providing direct access to every molecules dipole-moment vector. Contributions of slow dipolar cross-correlation have indeed been identified in several computer simulation studies. While such analyses are inherently limited to higher temperatures beyond the deeply supercooled regime, they confirm the experimental observations at low temperatures discussed above.

Koperwas et al.~\cite{koperwas2022computational} studied two versions of liquids with rhombus-like molecules, the only difference being the magnitude of the molecular dipole moments $\mu$. While for the low-$\mu$ system, the polarization autocorrelation function $C_P(t)$ almost exclusively contained contributions from self-correlations, it was dominated by cross-correlations in case of the high-$\mu$ system. These dipolar cross-correlations were confirmed to relax on significantly longer timescales than the self-correlations and were found to be Debye-shaped, thus confirming the experimental observations. Moreover, the choice of a model system with minimal additional types of intermolecular interactions allows to unambiguously attribute the dipolar cross-correlations to originating from dipole-dipole interactions.

A similar approach was taken by H\'{e}not et al. \cite{henot2023orientational} for the "real" liquid glycerol, by calculating self- and cross-correlations of the dipole-moment fluctuations determined  using standard MD-simulations. The authors identified significant cross-correlation contributions that decay slower than the self-correlations and with a larger high-frequency power law exponent. In addition, the geometry of these cross-correlations were analyzed, revealing a pattern of preferred orientations of adjacent dipoles that approximately reflect the dipole-dipole interaction potential, with contributions from additional effects, e.g. hydrogen-bonding or sterical constraints. Moreover, correlation functions associated with different Legendre polynomials of order $l=1$ and $l=2$ were calculated, thus mimicking a dielectric and a light scattering experiment, respectively. It turned out that the cross-correlations prominently present in $l=1$ are almost entirely absent in $l=2$. This result supports the empirical observation that light scattering is insensitive towards orientational cross-correlation effects. However, a recent MD study using a spherical tetrahedron-like molecule emphasizes the fact that cross-correlations in $l=2$ correlation functions must not be neglected in every case, as they found self- and cross-correlations of similar amplitude for this model molecule \cite{koperwas2024experimental}. Indeed, light-scattering experiments on certain monohydroxy-alcohols showed a weak additional slow process, possibly due to cross-correlations \cite{Gabriel2017}.

Several computational studies investigated the dynamics of liquid water and, just like in the works cited above, observed that its dielectric response is dominated by slow cross-correlations~\cite{Carlson2020,holzl2021dielectric,Alvarez2023}. Notably, one of these studies~\cite{holzl2021dielectric} used ab-initio molecular dynamics simulations (AIMD). In contrast to the classical MD simulations mentioned above, in AIMD simulations the electronic structure is explicitly taken into account, allowing for a realistic description of the molecular dipole moment under the influence of the surrounding molecules. This resulted in a dielectric spectrum in excellent agreement with experiment, highlighting the relevance of the finding of dominating cross-correlations in the dielectric spectrum of water. Since the dominant dielectric cross-correlation peak is absent in the light-scattering spectrum of water~\cite{fukasawa2005relation}, it would be instructive to calculate the light scattering spectrum based on the polarizability tensor which can be obtained from AIMD. In this way a more realistic comparison of dielectric and light scattering spectra could be performed compared to using the $l= 1, 2$ correlation function of sometimes rather arbitrarily chosen vectors as done in classical MD simulations.  

\section{Influence of intramolecular dynamics}
\label{sec:intra}

\begin{figure}[h]
   \includegraphics[width=0.48\textwidth]{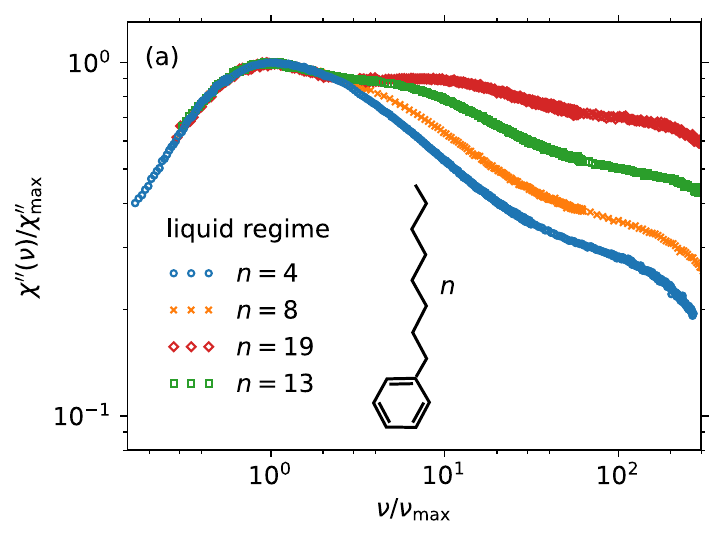}
   \includegraphics[width=0.48\textwidth]{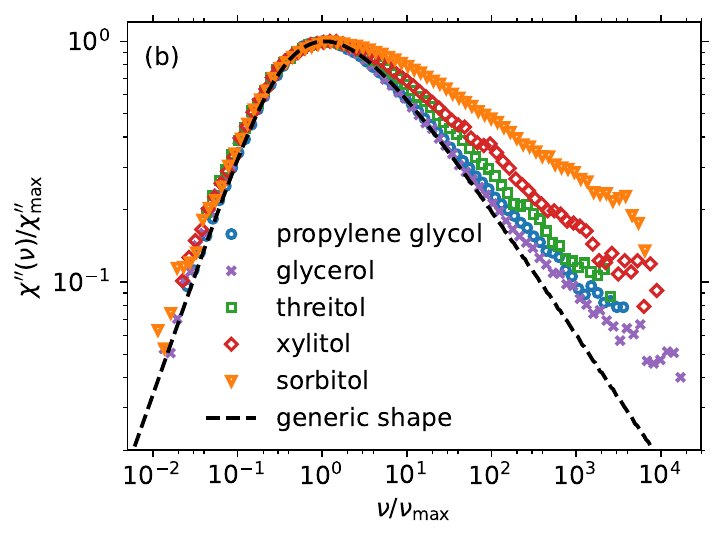}
   \caption{Examples of effects of intra-molecular dynamics on the spectral shape. (a) TFPI spectra of 1-phenylalkanes with varying number of carbons $n$ in the alkyl chain normalized to their maximum amplitude and slightly shifted in frequency to collapse their low frequency flanks. (b) DLS spectra of polyalcohols at temperatures close to the glass transition with varying number of OH-groups normalized to their respective peak maximum.} 
  \label{fig:ring-tail}
\end{figure}

The observation of a generic shape of structural relaxation for molecular glass formers with different chemical structures suggests that molecular details play a subordinate role in the deeply supercooled state. However, at temperatures far above the glass transition temperature, relaxation spectra among different substances were found to be much more diverse~\cite{schmidtke2014relaxation}. One can imagine several factors that could lead to this diversity, e.g., anisotropic reorientation \cite{hinze1995anisotropic}, local structure and dynamic asymmetry in, e.g., ionic liquids \cite{pabst2019mesoscale}, as well as the existence of intra-molecular degrees of freedom. 

Recently, Zeißler et al. \cite{zeissler2023influence} observed bimodal structural relaxation peaks for of 1-phenylalkanes above the melting point using high-frequency light-scattering techniques. Exemplary spectra of 1-phenylalkanes with different chain lengths are shown in \cref{fig:ring-tail}, each normalized with regard to peak-maximum amplitude and frequency. The bimodality becomes more pronounced with increasing chain-length and has been attributed to originate from dynamic decoupling of the ring group from the rest of the molecule, which gives rise to a fast contribution to the structural relaxation. Similar effects were observed in dielectric loss spectra of ring-tail structured molecules, were both, ring and tail carry a significant portion of the molecular dipole moment \cite{zeissler2023influence}, as well as in sizable glass formers with rigid non-polar cores attached to polar rotors \cite{rams2021complex}. Related effects of anisotropic molecular rotation were observed, e.g., for partially deuterated toluene in $^{2}$H-NMR experiments~\cite{hinze1995anisotropic}. 

Besides in liquids above the melting point, dynamic decoupling of different molecular moieties can also persist down into the deeply supercooled regime, as seen, e.g., for long-chained polyhydric alcohols (PAs). \cref{fig:ring-tail} shows DDLS spectra of supercooled PAs ranging from propylene glycol to sorbitol~\cite{bohmer2022glassy}. While short-chained PAs conform to the generic spectral shape, longer chained PAs show additional relaxation contributions at high-frequencies. Comparison to results from isotope labeled NMR experiments and to light-scattering results from above the melting points suggests that dynamic decoupling of different CH- and OH-groups within long-chained PA molecules could be the reason of the strong broadening of the structural relaxation peak~\cite{bohmer2022glassy,Doess2001,Sixou2001}. Such intramolecular dynamic heterogeneity leads to deviations from the generic high-frequency behavior. The reason for this effect being particularly pronounce in PAs is thought to be the hydrogen-bonded network, which prevents long-chained PA molecules to rotate as rigid entities, while this does not seem to be the case for short-chained PAs~\cite{Becher2021a}. Contrasting interpretations were given by Becher et al. \cite{Becher2021}, however, arguing that the differences of $^2$H-spin-lattice relaxation rates for different isotope labeled sites of sorbitol observed at high temperatures decreases upon cooling. Similar observations as for the PAs have been made for long-chained MAs in dielectric spectroscopy~\cite{Boehmer2023} and shear rheology~\cite{Arrese-Igor2020,Arrese-Igor2018}. Here, dynamic decoupling of the carbon backbone and the hydroxy group results in an intermediate relaxation contribution between the Debye- and the $\alpha$-relaxation. Similar to the broadening of the structural relaxation peak for the polyalcohols this interpretation is supported by NMR spectroscopy~\cite{Schildmann2011}, where rotational relaxation times of the OH- or OD-bonds was found to agree well with timescale of the intermediate process.

To summarize, the influence of molecular details and internal degrees of freedom on the spectral shape of structural relaxation can not be simply disregarded, especially in the liquid regime above the melting point. While such effects seem to be less relevant at supercooled temperatures~\cite{Becher2021}, they can play a role in certain cases, e.g., in longer-chained hydrogen-bonding liquids.

\section{Perspectives and open questions}
\label{sec:perspectives}

\subsection{What is special about $\nu^{-1/2}$?}

The observation of a generic $\nu^{-1/2}$ high-frequency power law in a broad variety of deeply supercooled liquids is striking and suggests a deeply rooted connection to the glass-transition phenomenon. An obvious question is: What is special about $\nu^{-1/2}$? And what is the underlying physical origin of this exact power law behavior? Interestingly, numerous theoretical studies following different approaches have predicted the $\nu^{-1/2}$ high-frequency behavior (see Ref.~\cite{Dyre2005} and references therein for a brief review). 

Recently, Dyre provided a review of the \textit{solid-that-flows} perspective on deeply supercooled liquids, which involved a qualitative discussion of the structural spectral shape~\cite{Dyre2024}. It was argued that structural relaxation proceeds via various localized flow events that start in highly mobile regions with low local energy barriers and finally induce dynamic facilitation by reducing large local energy barriers in less mobile regions. The time scale at which dynamic facilitation mobilizes even the least mobile regions marks the long-time cutoff of the distribution of relaxation times and, thus, determines the frequency below which the $\nu^1$ low-frequency behavior is observed. At larger frequencies, solidity leads to the asymmetric loss peak with a $\nu^{-\beta}$ high-frequency behavior. In previous considerations~\cite{Dyre2006}, $\beta=1/2$ was predicted by translating the solid-that-flows conjecture into an expression for density fluctuations, which was then solved in the Gaussian approximation.

Dynamic facilitation in supercooled liquids has also been studied in recent computer-simulation approaches~\cite{Scalliet2022,Guiselin2022}. In harmony with the above mentioned notions it was found that relaxation begins in localized highly-mobile regions that facilitate relaxation events in their vicinity. Interestingly, close connections were identified between dynamic facilitation and the high-frequency power-law exponent of the respective relaxation spectra: The high-frequency power-law exponent extracted from the relaxation spectra was found to also describe the short-time evolution of the number of highly-mobile clusters, as well as the waiting time distribution for the emergence of new clusters.\cite{Scalliet2022}

Considering these findings, it is tempting to contemplate whether the generic high-frequency power law might reflect generic features of dynamic facilitation. Computer simulations are a promising tool to elucidate this conjecture. However, drawing more quantitative conclusions for molecular liquids will require models that more closely resemble molecular liquids, e.g. by reducing the polydispersity~\cite{Pihlajamaa2023} or implementing rotational degrees of freedom~\cite{Ozawa2023,Ozawa2024}.

\subsection{How relevant are cross-correlations for the glass-transition phenomenon?}
While the presented results suggest a generic spectral shape for self-correlations, the properties of dipolar cross-correlations seem to sensitively depend on molecular peculiarities. It remains an open question to what degree the relaxation of these cross-correlations is related to the glass transition phenomenon. In particular, the question arises as to whether the transition from an equilibrium supercooled liquid to an off-equilibrium glass during cooling, and vice versa, is controlled by the characteristic time-scale of self-correlations or that of cross-correlations. Promising approaches to tackle this question involve the study of physical aging, either during cooling and subsequent heating with fixed rates, or after performing instantaneous temperature-step protocols and monitoring the transition from one to the other equilibrium state, for instance by using a recently introduced experimental approach using light-scattering~\cite{Boehmer2024}. For monohydroxy alcohols it is well established that the glass transition temperature and physical aging couple to the characteristic time scale of self-correlations~\cite{Hansen1997,Huth2007,Gainaru2010a,Preuss2012,Boehmer2014}. Whether the same holds for other classes of supercooled systems like polyhdric alcohols or polar van-der-Waals liquids remains an open question. First experiments on tributyl phosphate and propylene glycol suggest that in these systems the glass transition might instead be coupled to the relaxation time of cross-correlations~\cite{Moch2022, Richert2022}. 

\subsection{Why does DDLS not resolve cross-correlations in supercooled liquids?}
As discussed above, it was found empirically that DDLS is largely insensitive to orientational cross-correlations and only probes the self-correlation contributions to the relaxation spectrum. By contrast, the theoretical treatment of light scattering yields a different picture: Similar to what was discussed above for the dielectric properties, the electric field time-autocorrelation function probed by DDLS can be approximated as
\begin{equation}
    C_{\bm{E}}(t) = g_1(t) \propto \sum\limits_i 
 \bigl\langle P_2[\bm{u}_i(0)\cdot\bm{u}_i(t)]\bigr\rangle  + \sum\limits_{i\neq j} \bigl\langle P_2[\bm{u}_i(0)\cdot\bm{u}_j(t)]\bigr\rangle,
 \label{equ:crossDDLS}
\end{equation}
where $\bm{u}$ denotes the vector along the principle axis of the molecular polarizability tensor and $P_2(x)=(3x^2+1)/2$ is the second order Legendre polynomial. By contrast, the dot product in the treatment of dielectric relaxation \cref{equ:selfcross} corresponds to the first order Legendre polynomial $P_1(x)=x$. As obvious from \cref{equ:crossDDLS}, orientational cross-correlations can, in principle, contribute to the electric field time-autocorrelation function. The difference to dielectric relaxation is, however, the corresponding angular sensitivity, e.g., in case of a $P_1$ correlation function negative orientational cross-correlations correspond to an angle of 180$^\circ$, but to 90$^\circ$ for a $P_2$ correlation function. These different angular sensitivities might explain the empirical observations that cross-correlations contribute strongly to the dielectric loss, but are negligible in DDLS, while the dynamics of the self-correlations are comparable between both techniques. This conjecture has recently been confirmed for glycerol by H\'enot et al.\ using computer simulations~\cite{henot2023orientational}. The authors showed that cross-correlations contribute to the $P_1$ correlation function as a slow process, while they are orders of magnitude weaker in $P_2$. Generalizing these findings to different types of molecular liquids should be an aim of future research.

\subsection{Transition between low- and high-temperature regime}
\begin{figure}[h]
   \includegraphics[width=0.48\textwidth]{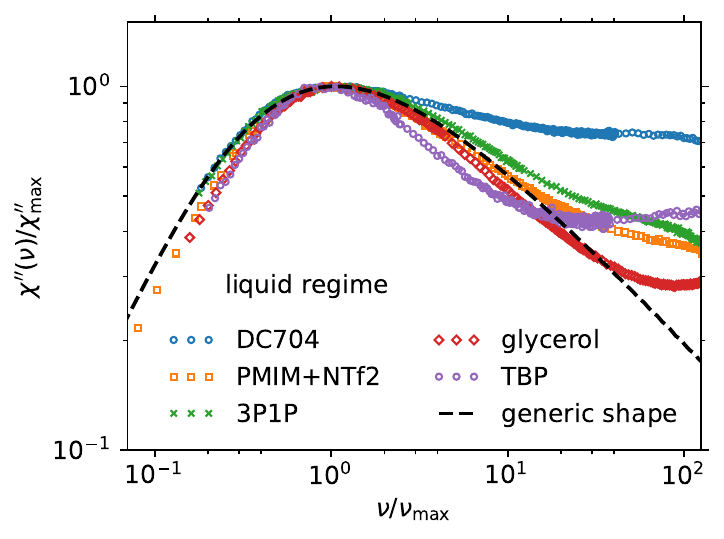}
   \caption{High frequency DDLS spectra (\SI{}{\mega\hertz} to \SI{}{\tera\hertz}) of several liquids obtained with a tandem Fabry-Perot interferometer at temperatures above the melting point. A variety of spectral shapes is observed, including more and less pronounced relaxation stretching compared to the deeply supercooled regime.} 
  \label{fig:high-temperature}
\end{figure}

Another perspective concerns the temperature evolution of the spectral shape of structural relaxation going from the deeply supercooled to the liquid state. A common notion has been that the asymmetry of the relaxation peak decreases upon increasing temperature, reflecting dynamic heterogeneity becoming less pronounced, until, ultimately, the spectral shape converges towards an exponential far above the melting point \cite{lunkenheimer2000glassy,Schoenhals1993a,Dixon1994comment}. Results from more recent work seems to contradict these conjectures, as illustrated exemplarily in \cref{fig:high-temperature}, showing high-frequency light-scattering data of different liquids at relaxation times of $\sim\SI{50}{\pico\second}$. All of the included liquids were shown to have a very similar spectral shape with $\nu^{-1/2}$ high-frequency behavior at deeply supercooled temperatures (see \cref{fig:master}b). By contrast, much more diverse spectral shapes are found in the liquid regime, which, however, are not always more narrow than the spectral shape at deeply supercooled temperatures. Instead, the high-temperature shape is found to be more broad, similar, and less broad than at deeply supercooled temperatures for the shown examples. Similar results were obtained by Schmidtke et al. \cite{schmidtke2014relaxation} in a study investigating a number of molecular liquids and room temperature ionic liquids employing mainly high-frequency light-scattering experiments in the \SI{}{\mega\hertz} to \SI{}{\tera\hertz} regime. Not only did this study reveal that the spectral shape of structural relaxation is rather diverse in the mildly supercooled and liquid regime, i.e., stretching parameters of the Cole-Davidson model ranging from $0.3$ to $0.8$ were reported, but in addition many of the liquids were found to obey FTS in the liquid regime. Recently, Rössler et al. \cite{Roessler2024} showed in a detailed comparison of light scattering, dielectric and NMR data for several glass formers that the observation of FTS in the \SI{}{\mega\hertz} to \SI{}{\tera\hertz} regime is not exclusive to DDLS, but is confirmed by NMR relaxometry. 

Considering these results, it seems likely that there exist two regimes obeying FTS, one in the deeply supercooled and one in the liquid state. Between both regimes, a transition between low- and high-temperature shape seems to occur - at least in those cases where the two are not identical (see example in \cref{fig:tts}). Performing experiments to test this hypothesis and to characterize the transition should be subject of future work.

\subsection{What about polymers?}

Besides low molecular-weight substances, polymers are another important class of materials that commonly are obtained in the deeply supercooled liquid state. Yet, we did not discuss any aspects of the spectral shape of structural relaxation in polymers within this work. The reason for this mainly is that, as stated by Dyre, "Polymers are different."~\cite{Dyre2007} In contrast to low molecular-weight supercooled liquids, supercooled polymers do not flow above the glass transition temperature as a result of chain-connectivity effects like entanglement~\cite{Strobl1997}. Thus, relaxation in polymers is comprised of a complex interplay between the relaxation of units of different sizes, ranging from single conformers, i.e., units involved in local conformational changes, over segments consisting of a small number of repeat units to, lastly, the whole chain (Rouse modes). Indeed, Baker et al. recently presented convincing evidence that the structural relaxation in polymers is incorporated in a hierarchical scheme of relaxations starting with the relaxation of single conformers, i.e. dihedral rotations~\cite{Baker2022}. These authors suggest a dynamic facilitation mechanism that can be of inter- or intra-molecular nature depending on the molecular weight of the polymer chains. This suggests that, at least to a certain extent, structural relaxation in polymers reflects more local properties, e.g., chemical structure, as well as local environments of conformers. In fact, structural relaxation in polymers was found to feature intrinsic relaxation stretching~\cite{Heuer1999,Richert2002,DiazVela2020}, in the sense that it can not be described in terms of a superposition of exponentials with distributed relaxation times. The relaxation stretching observed for macroscopic variables is thus comprised of stretching due to spatial dynamic heterogeneity, as well as intrinsic stretching, at least partly, due to chain connectivity~\cite{DiazVela2020}. 

Thus, we argue that the spectral shape of structural relaxation in polymers should be discussed separately from that of low molecular-weight supercooled liquids. Therefore, it is not to be expected that structural relaxation in polymers conforms to the generic $\nu^{-1/2}$ high-frequency behavior.

Indeed, previous experimental work on polymeric supercooled liquids using light-scattering or dielectric spectroscopy showed the spectral shape of structural relaxation to be more diverse, commonly reporting values of $\beta~\sim 0.3-0.5$~\cite{Lee_1979,Boese_1989,Fytas_1985,Fytas_1988,Patterson_1979,Patterson1983,Ngai_1988,Wang1981}. It has to be noted though that these values were mostly obtained from curve-fitting procedures, commonly applied using linear instead of logarithmic residuals. Another interesting aspect is the molecular-weight dependence of the spectral shape. First results in this regard were obtained by Hintermeyer et al. \cite{Hintermeyer2000}. From dielectric-loss data these authors found the spectral shape of structural relaxation to be virtually independent of molecular weight for PDMS, while PS showed a discontinuous change of the spectral shape at molecular weights associated with about one Kuhn length. Considering these points we suggest to revisit the spectral shape of structural relaxation in different supercooled polymers taking into account the aspects discussed within this article and searching for generic features and correlations with chemical structure. 

\subsection{Conclusions}
In this Perspective, we discussed the present state of evidence concerning a generic spectral shape in the reorientational susceptibility spectra of deeply supercooled molecular liquids, in particular from dielectric and depolarized light scattering spectroscopy. The most prominent generic feature seems to be the $\nu^{-1/2}$ high frequency power law in the spectrum of self-correlations observed in light scattering in a large variety of different systems including van der Waals, hydrogen bonding and even ionic liquids. In the dielectric loss, the generic shape is most often superimposed by cross-correlation contributions and is therefore directly evident only in molecular liquids with low dipole moment while with increasing polarity and in particular increasing Kirkwood correlation $g_K$ deviations due to cross-correlations become visible. Such cross terms can be suppressed, e.g., by dilution, pressurization, or hyperquenching of the liquid. In addition to cross-correlations, deviations from generic behavior also occur due to intramolecular degrees of freedom and most likely also in the cases of very anisotropic molecules and of chain connectivity in polymers.

Thus, all of these findings point to a generic aspect of supercooled liquid dynamics, which becomes evident in simple molecular liquids and so far remains poorly understood. In  particular the relation with dynamic heterogeneity and dynamic facilitation and the special role of the $\nu^{-1/2}$ power law in the deeply supercooled state needs to be further explored. Also, correlations of the spectral shape with other aspects of supercooled liquid dynamics, many of which were suggested in the past, need careful reconsideration, as in most of these previous concepts the crucial role of cross-correlations for the diversity of spectral shapes was not considered. One of the benefits of these findings seems to be that, by comparing light scattering and dielectric spectra, for many liquids the self- and cross-correlation contributions can, at least approximately, be separated. Still, it requires further clarification in how far dipolar cross-correlations play an important role for the relaxation of the overall structure in the supercooled state, and, most likely, different types of cross-correlations need to be distinguished in that respect. Finally, also the fact that DDLS seems to be largely unaffected by cross correlations is not completely clarified yet, and, like the other questions posed above, will benefit not only from further developments of theory but also from molecular dynamics simulations, in particular those that extend into the deeply supercooled state. 

\subsection{Acknowledgments}
The authors thank Pierre-Michel Déjardin, Jeppe C. Dyre, Ranko Richert, Ernst Rößler, Marian Paluch, Roland Böhmer, Catalin Gainaru and Silvia Arrese-Igor for fruitful discussions, as well as Ranko Richert, Ernst Rößler and Marian Paluch for kindly providing experimental data. Financial support by the Deutsche Forschungsgemeinschaft under grant no. BL 1192/1 and 1192/3 is gratefully acknowledged.

\bibliographystyle{achemso} 
\bibliography{relax-shape}
\end{document}